\documentclass[reprint, prl, aps, showpacs]{revtex4-1}

\usepackage{graphicx}
\usepackage{amsmath,amssymb}
\usepackage{epstopdf}
\usepackage[usenames, dvipsnames]{color}
\usepackage{subfig}
\usepackage{verbatim}
\usepackage{caption}
\usepackage[utf8]{inputenc}
\usepackage{tikz}
\usetikzlibrary{shapes,arrows}

\newcommand{\beq}{\begin{equation}}
\newcommand{\eeq}{\end{equation}}
\newcommand{\bdm}{\begin{displaymath}}
\newcommand{\edm}{\end{displaymath}}

\DeclareFontFamily{OT1}{pzc}{}
\DeclareFontShape{OT1}{pzc}{m}{it}{<-> s * [1.10] pzcmi7t}{}
\DeclareMathAlphabet{\mathpzc}{OT1}{pzc}{m}{it}

\begin{document}

\title{Control strategy to limit duty cycle impact of earthquakes on the LIGO gravitational-wave detectors.}

\author{S. Biscans\textsuperscript{1}}\thanks{Corresponding author: sbiscans@ligo.mit.edu}
\author{J. Warner\textsuperscript{2}}
\author{R. Mittleman\textsuperscript{1}}
\author{C. Buchanan\textsuperscript{3}}
\author{M. Coughlin\textsuperscript{4}}
\author{M. Evans\textsuperscript{1}}
\author{H. Gabbard\textsuperscript{5}}
\author{J. Harms\textsuperscript{6}}
\author{B. Lantz\textsuperscript{7}}
\author{N. Mukund\textsuperscript{8}}
\author{A. Pele\textsuperscript{9}}
\author{C. Pezerat\textsuperscript{10}}
\author{P. Picart\textsuperscript{10}}
\author{H. Radkins\textsuperscript{2}}
\author{T. Shaffer\textsuperscript{2}}

\affiliation{\textsuperscript{1}LIGO, Massachusetts Institute of Technology, Cambridge, Massachusetts 02139, USA}
\affiliation{\textsuperscript{2}LIGO Hanford Observatory, Richland, Washington 99352, USA}
\affiliation{\textsuperscript{3}Department of Physics and Astronomy, Louisiana State University, Baton Rouge, LA 70803, USA}
\affiliation{\textsuperscript{4}Department of Physics, Harvard University, Cambridge, MA 02138, USA}
\affiliation{\textsuperscript{5}Albert-Einstein-Institut, Max-Planck-Institut f\"{u}r Gravitationsphysik, D-30167 Hannover, Germany}
\affiliation{\textsuperscript{6}Universit\`a degli Studi di Urbino ``Carlo Bo'', I-61029 Urbino, Italy}
\affiliation{\textsuperscript{7}Stanford University, Stanford, California 94305, USA}
\affiliation{\textsuperscript{8}Inter-University Centre for Astronomy and Astrophysics (IUCAA),\\ Post Bag 4, Ganeshkhind, Pune 411 007, India}
\affiliation{\textsuperscript{9}LIGO Livingston Observatory, Livingston, Louisiana 70754, USA}
\affiliation{\textsuperscript{10}Université du Maine, CNRS UMR 6613, 72085 Le Mans, France}

\begin{abstract}

Advanced gravitational-wave detectors such as the Laser Interferometer Gravitational-Wave Observatories (LIGO) require an unprecedented level of isolation from the ground. When in operation, they are expected to observe changes in the space-time continuum of less than one thousandth of the diameter of a proton. Strong teleseismic events like earthquakes disrupt the proper functioning of the detectors, and result in a loss of data until the detectors can be returned to their operating states. An earthquake early-warning system, as well as a prediction model have been developed to help understand the impact of earthquakes on LIGO. This paper describes a control strategy to use this early-warning system to reduce the LIGO downtime by $\rm \sim$30\%. It also presents a plan to implement this new earthquake configuration in the LIGO automation system.

\end{abstract}

\maketitle
\section{Introduction}

The Laser Interferometer Gravitational-wave Observatory (LIGO) consists of two identical, 4 kilometers long interferometric detectors installed at the Hanford, Washington (H1) and Livingston, Louisiana (L1) sites in the United States. The detectors are Michelson interferometers with Fabry-Perot resonant cavity arms \cite{abbott2016gw150914}. A prerequisite for the detectors to collect scientific data is that the cavities are held in optical resonance.

LIGO completed its most recent observation run (referred as O1) from September 18, 2015 to January 12, 2016. During this period, the detectors were kept in an operating mode and the commissioning activities reduced to minimum. Keeping the detectors in operating mode is a complex task and large environmental disturbances can disrupt this process. These disturbances reduce the duty cycle and observing time, as summarized in table \ref{table:causes}.

\begin{figure}[t]
\hspace*{-0.9cm}
\centering
\includegraphics[width=4in]{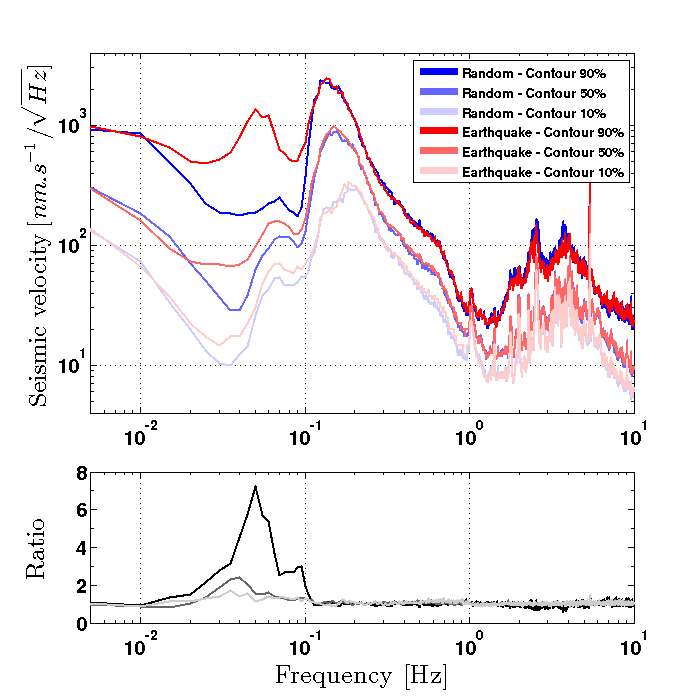}
\caption{1000-second long data stretches were selected over the total time span of the O1 period at H1. Specifically, the blue curves correspond to stretches selected at random times, while the red curves correspond to stretches selected during Earthquakes of Richter magnitude 4.5 and above.  For each frequency bin, the data were histogrammed and a set of probabilities was found. The different shades of colour indicate different probability values (10\%, 50\%, 90\%), the darker tone being a 90\% probability. The bottom part of this figure represents the ratio between the red and the blue curves for each set. We observe an amplification of the ground motion up to a factor of 7 in the [30mHz-100mHz] bandwidth. Only the horizontal direction along the Y-arm of the interferometer is represented here, but we obtain similar results in the X-horizontal direction and the vertical direction.}
 \label{fig:spectra}
\end{figure}

Environmental hazards such as earthquakes were one of the primary sources of disturbance during the first observation run. Earthquakes with a Richter magnitude above 4.5 generated a non-negligible increase of the ground motion in vertical and horizontal directions, from a few mHz to 100mHz, as illustrated by the seismic vibration spectra in figure \ref{fig:spectra}. This increase of the ground motion overwhelmed the LIGO seismic isolation system and prevented the detectors from operating. Once the cavities of the interferometer are out of resonance, it can take many hours to return to normal operation, as demonstrated in figure \ref{fig:timelock}. Overall, H1 experienced 265 earthquakes and L1 243 earthquakes (magnitude 4.5 and above) while observing. Operation was disrupted 62 times at Hanford and 83 times at Livingston due to these earthquakes. This study will only focus on Hanford, knowing that similar results are expected for Livingston. 

\begin{figure*}[t]
\hspace*{-0.5cm}
    \centering
    \subfloat[Ground]{{\includegraphics[width=3.25in]{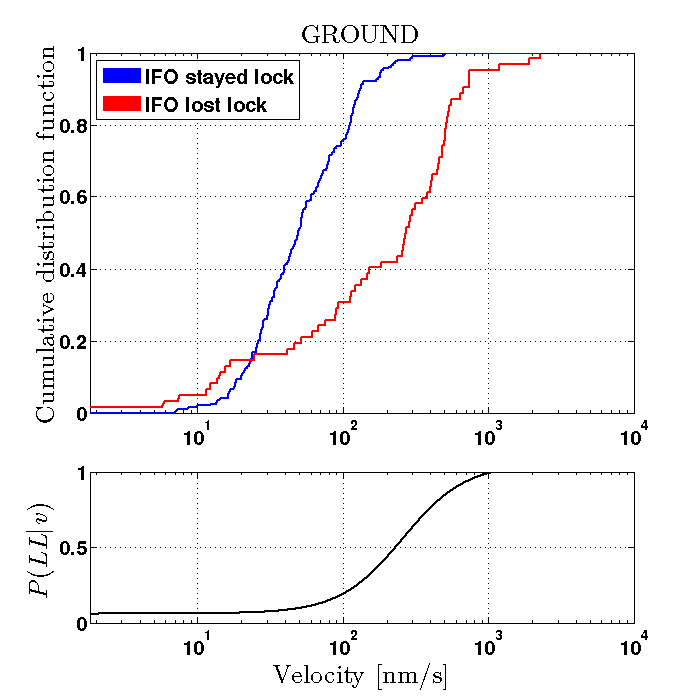} }}%
    \qquad
    \subfloat[Stage]{{\includegraphics[width=3.25in]{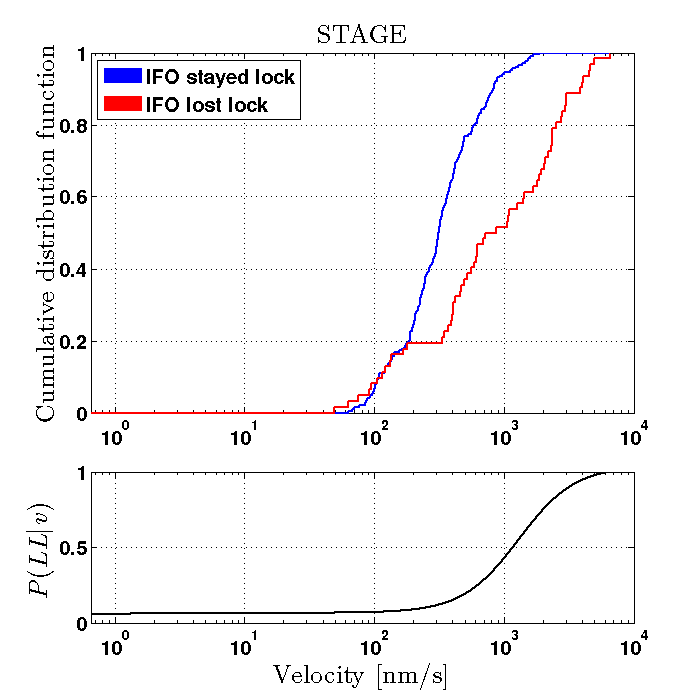} }}%
    \caption{Comparison of Stage 1 ITM behavior in the [30mHz-100mHz] bandwidth for different ground motions: stretches selected during earthquakes when the interferometer survived (blue curve), stretches selected during earthquakes when the interferometer stops functioning (red curve). The top part of the figure represents the cumulative distribution function for the ground and the stage respectively, as a function of velocity. The plots indicate a direct correlation between velocity and the interferometer status. We observe a net increase of the stage velocity compared to the ground, due to a self-inflicted gain peaking in this frequency band. The bottom part of the plots represents the smoothed probability of losing lock (dropping out of optical resonance) as a function of velocity.}
 \label{fig:bins}
\end{figure*}

There is a direct correlation between the operating status of the interferometers and the ground velocity, as shown in figure \ref{fig:bins}a: the interferometer becomes unstable at higher velocities. An Earthquake mitigation scheme is thus developed to limit the extra disturbance induced by earthquakes. The goal of this configuration is to maintain operation at the expense of sensitivity. For this reason, it cannot be permanently activated and needs to be part of a smart automation system.

\begin{table}[h!]
\begin{center}
\begin{tabular}{l*{6}{c}r}
              & H1 & L1 \\
\hline
Observation & 66 \% & 59 \% \\
Commissioning & 9 \% & 7 \% \\
\textbf{Environmental disturbances} & \textbf{17 \%} & \textbf{24 \%} \\
Other & 8 \% & 10 \% \\
\end{tabular}
\end{center}
\caption{Detectors' status over the O1 period. Commissioning time represents the vital maintenance tasks needed to keep the interferometers running. Environmental disturbances encompass earthquakes, high-wind and distant storms.}
\label{table:causes}
\end{table}

The first section of this paper describes the LIGO interferometers and its seismic isolation systems in more detail. We then present the earthquake strategy and how it will be integrated into the LIGO infrastructure.

\section{The LIGO detectors}

\begin{figure}[t]
\hspace*{-0.5cm}
\centering
\includegraphics[width=4in]{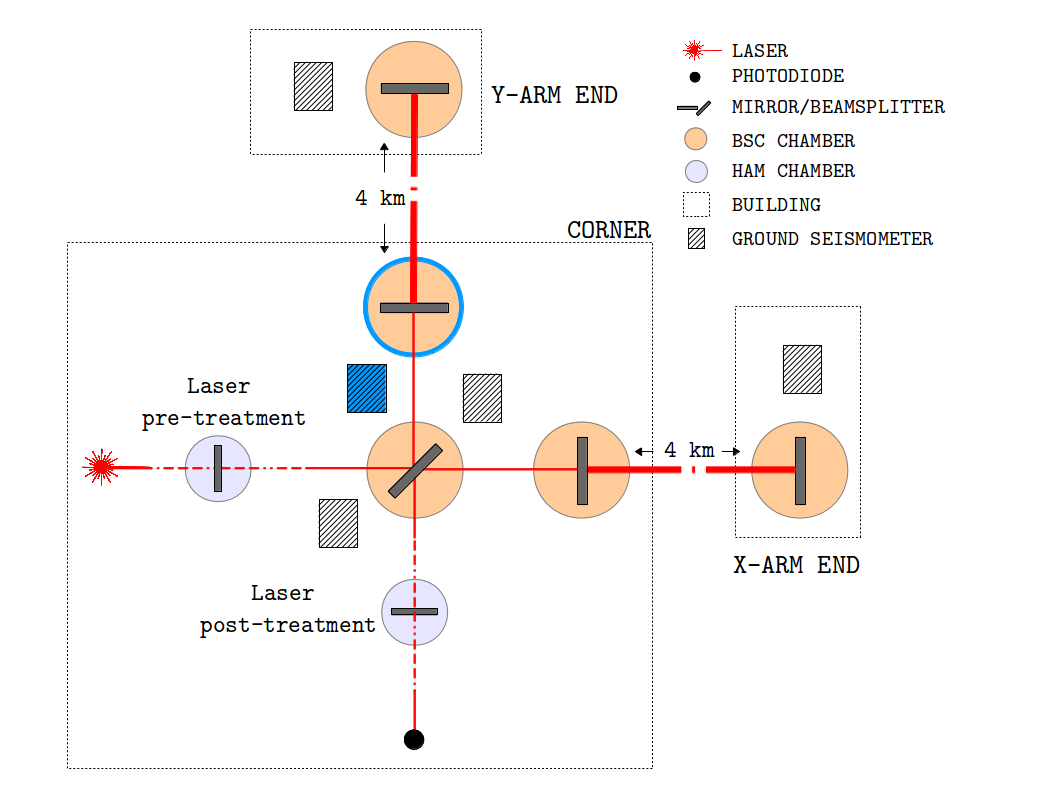}
\caption{Simplified optical layout of the LIGO detector. The data presented in this study is extracted from the blue ground seismometer and the BSC chamber circled in blue (called ITMY).}
 \label{fig:layout}
\end{figure}

The LIGO detectors are Michelson interferometers coupled with Fabry-Perot cavities in the arms. A beamsplitter is used to separate the input light into the two arms. A simplified optical layout of the LIGO detector is shown in figure \ref{fig:layout}. Each arm is comprised of an input test mass (ITM) and an output test mass (ETM) forming the Fabry-Perot arm cavity. Other cavities and auxiliary optics such as the power recycling mirror and the signal recycling mirror are present to enhance the signal quality \cite{aasi2015advanced}. 

All the LIGO's optics are attached to seismic isolation platforms, which seek to decouple the optics from the ground in the observational bandwidth (above 10Hz). A Hydraulic External Pre-Isolator (HEPI) supports the vacuum chamber. The optics are mounted inside the chambers on an active Internal Seismic Isolation (ISI) system. There are two types of ISI systems: a Single-Stage platform and a Two-Stages platform. The Single-Stage platforms are used for the auxiliary optics in the small vacuum chambers called HAMs. The Two-Stages platforms are used for the core optics of the interferometers, in the large vacuum tanks called BSCs.  In total, a detector has 11 vacuum tanks (six HAM chambers, five BSC chambers) each with a seismic isolation platform. Despite some mechanical differences between the different vibration isolation systems, the general concept is identical for all of them. A combination of active and passive isolation is provided to bring the BSC-ISI platform motion down to $\rm 1\cdot10^{-12} \ m/\sqrt{Hz}$ at 10Hz and the HAM-ISI platform motion down to $\rm 2\cdot10^{-11} \ m/\sqrt{Hz}$ at 10Hz.  

\begin{figure}[t]
\hspace*{-0.5cm}
\centering
\includegraphics[width=3in]{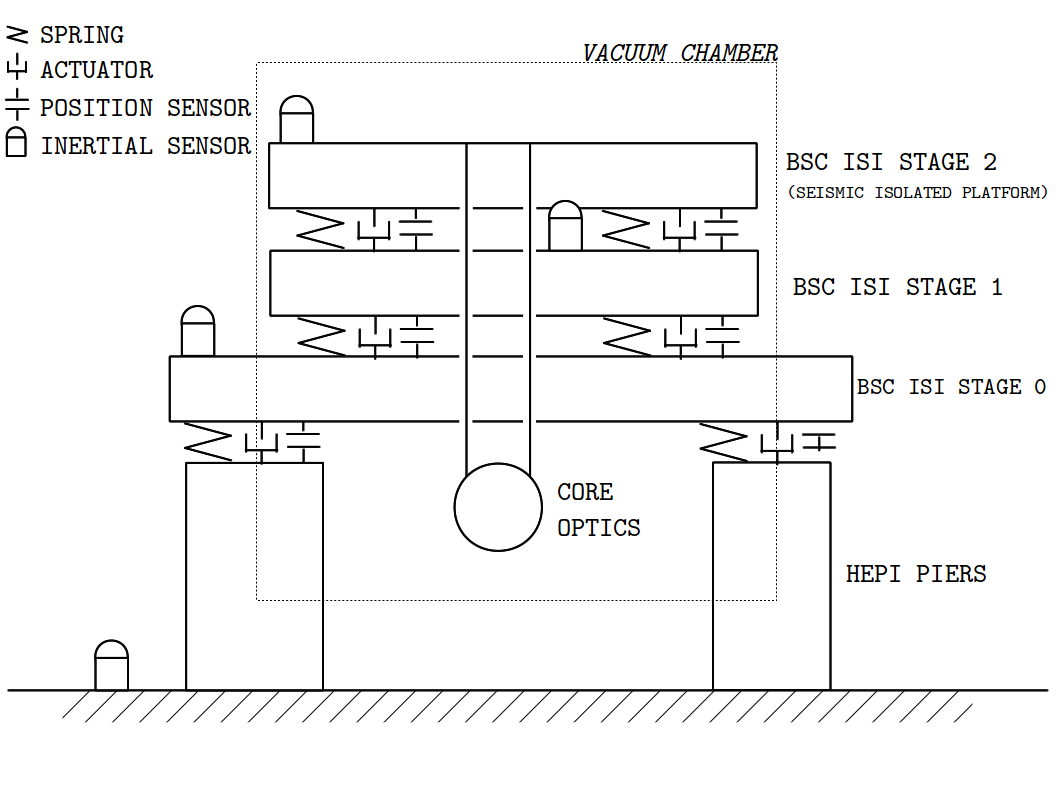}
\caption{Schematic of the LIGO BSC chamber. Each stage is equipped with multiple actuators, position and inertial sensors (only a few are represented here for clarity). The core optics are supported by a quadruple pendulum (not shown for clarity) which provides additional seismic isolation in all degrees of freedom.}
 \label{fig:vacuum}
\end{figure}

The mechanics and functioning of these platforms have been extensively studied in previous works (\cite{matichard2015seismic}, \cite{abbott2002seismic}, \cite{abbott2004seismic}). This paper will focus only on the active isolation configuration and performance of the BSC-ISI Stage 1 platform. The BSC-ISI Stage 1 platform is of interest for this study as it provides low frequency active isolation for the core optics. All the BSC-ISI platforms being identical, we chose to look at the ITM chamber of the Y arm (called ITMY) in the horizontal Y-direction. Figure \ref{fig:vacuum} illustrates the layout of the BSC chamber.

\section{The Seismic Platform Control scheme}
\label{sec:control}

Each stage is equipped with a set of actuators, displacement sensors and inertial sensors. They are used to actively control the stage in the three translational and three rotational degrees of freedom. The platforms have been designed to limit the cross-coupling between the different degrees of freedom, therefore, each degree of freedom can be actively controlled independently with Single-Input Single-Output compensators.

\begin{figure}[t]
\hspace*{-0.3cm}
\centering
\includegraphics[width=3.5in]{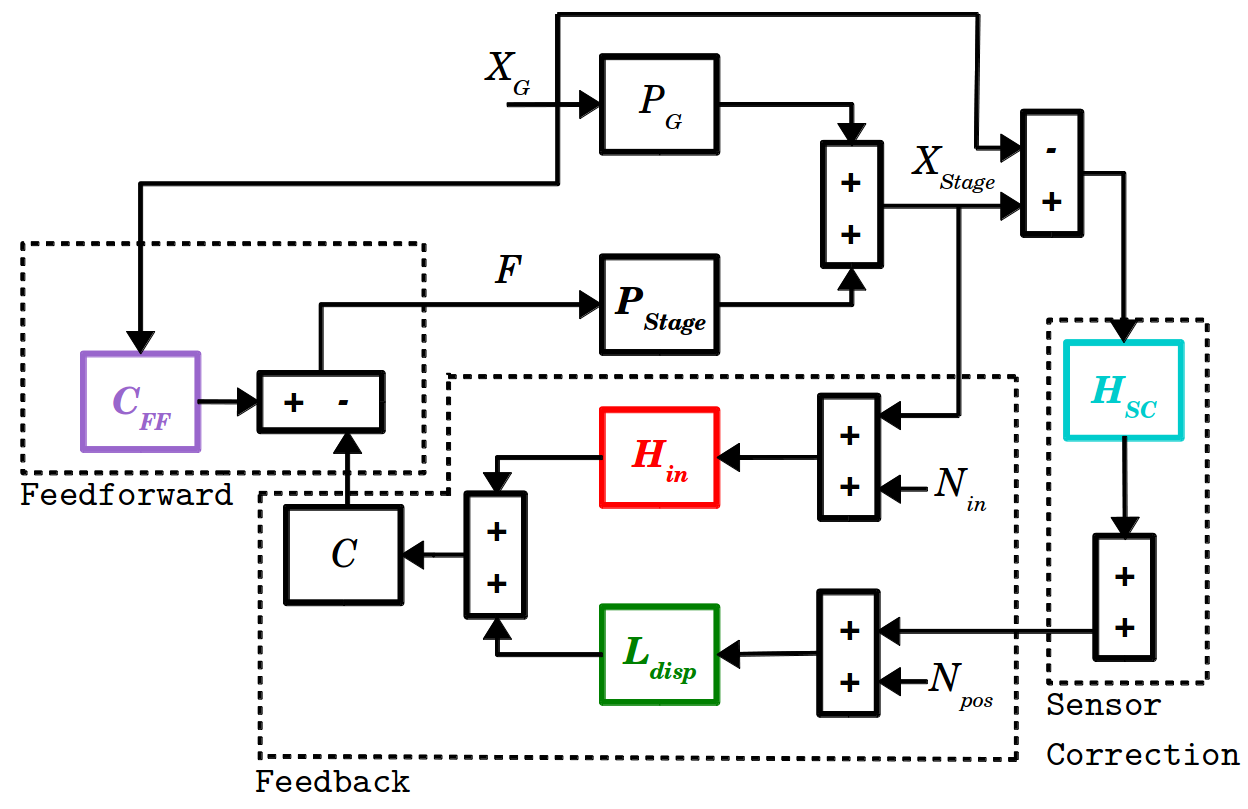}
\caption{Control block diagram of a LIGO seismic isolation stage for one degree of freedom. The coloured blocks are related to figure \ref{fig:closeloop}.}
 \label{fig:control}
\end{figure}

The LIGO seismic control scheme is a combination of feedback, feedforward and sensor correction. The block diagram in figure \ref{fig:control} shows the simplified control topology for one degree of freedom.

\subsubsection{Feedback control}

A control force $\rm F$ is used to reduce the inertial motion of the stage ($\rm X_{Stage}$), which is induced by the ground ($\rm X_{G}$). This control force is created using a combination of feedback and feedforward controllers. The feedback controller is fed by two sensors: a displacement sensor measuring the relative motion between the stage and the input motion ($\rm X_{Stage}-X_{G}$), and an inertial sensor (seismometer) measuring the inertial motion of the stage ($\rm X_{Stage}$). We wish to reduce the inertial motion measured by the seismometers, but cannot below 25mHz because of noise limitations. Therefore, a displacement sensor is used at low frequencies and both sensors are blended together to feed the controller. The relative motion signal is low-passed by a filter $\rm L_{disp}$, and the inertial motion signal is high-passed by a filter $\rm H_{in}$. $\rm L_{disp}$ and $\rm H_{in}$ are designed to be complementary, meaning $\rm L_{disp}+H_{in} = 1$. The frequency at which the low-pass and high-pass filters cross is called the blend frequency. The controller provides isolation up to 30Hz, with high loop gain below 1Hz. The expression of the stage inertial motion when the feedback control is on and the feedforward and sensor correction are off is given by equation \ref{eq:fb}. $\rm N_{pos}$ and $\rm N_{in}$ represent the noise associated with the displacement sensors and inertial sensors respectively. $\rm P_{G}$ and $\rm P_{Stage}$ represent the plant model of the ground and stage respectively.

\begin{equation}
\begin{aligned}
|X_{Stage}|^{2}={} & |\frac{P_{G}+L_{disp}CP_{Stage}}{1+CP_{Stage}}X_{G}|^{2} \\
      & + |\frac{H_{in}CP_{Stage}}{1+CP_{Stage}}N_{in}|^{2} \\
      & + |\frac{L_{disp}CP_{Stage}}{1+CP_{Stage}}N_{pos}|^{2}
\end{aligned}
\label{eq:fb}
\end{equation}

\subsubsection{Sensor Correction}
Sensor correction is a feedforward technique using a seismometer from the ground. The seismometer signal $\rm X_{G}$ is filtered by $\rm H_{SC}$ and added to the position sensor signal ($\rm X_{Stage}-X_{G}$) to create a virtual inertial sensor \cite{senscor}. Our sensor correction is designed to increase performance around 100mHz. By adding the sensor correction to the feedback loop, the stage inertial motion becomes:

\begin{equation}
\begin{aligned}
|X_{Stage}|^{2}={} & |\frac{P_{G}+L_{disp}CP_{Stage}(1-H_{SC})}{1+CP_{Stage}}X_{G}|^{2} \\
      & + |\frac{H_{in}CP_{Stage}}{1+CP_{Stage}}N_{in}|^{2} \\
      & + |\frac{L_{disp}CP_{Stage}}{1+CP_{Stage}}N_{pos}|^{2}
\end{aligned}
\label{eq:sc}
\end{equation}

\subsubsection{Feedforward control}
A standard feedforward controller is added from the ground in parallel with the feedback and sensor correction loops. This operates where the coherence with the ground is high (above 1Hz). The stage absolute motion becomes:

\begin{equation}
\begin{aligned}
|X_{Stage}|^{2}={} & |\frac{P_{G}+L_{disp}CP_{Stage}(1-H_{SC})+C_{FF}P_{Stage}}{1+CP_{Stage}}X_{G}|^{2} \\
      & + |\frac{H_{in}CP_{Stage}}{1+CP_{Stage}}N_{in}|^{2} \\
      & + |\frac{L_{disp}CP_{Stage}}{1+CP_{Stage}}N_{pos}|^{2}
\end{aligned}
\label{eq:sc2}
\end{equation}

\begin{figure}[t]
\hspace*{-0.75cm}
\centering
\includegraphics[width=3.5in]{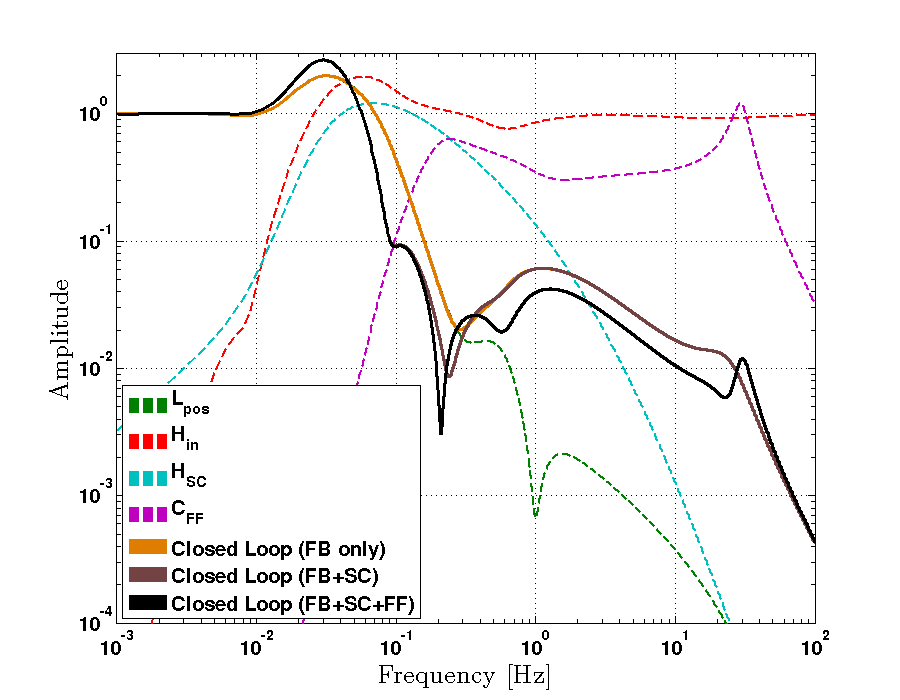}
\caption{Example of the LIGO seismic control scheme on a damped single degree of freedom system. The dash curves shows the different filters used, as opposed to the solid curves showing the transfer functions with different loops engaged.}
 \label{fig:closeloop}
\end{figure}

Equations \ref{eq:fb}, \ref{eq:sc} and \ref{eq:sc2} are plotted in an example in figure \ref{fig:closeloop} to illustrate the performance of each loop and the combined overall performance. In this example, we used the nominal BSC-ISI stage 1 filters on a damped single degree of freedom system with a 2Hz resonance (which is similar to the BSC-ISI Stage 1). The figure shows the transfer function between the stage and ground motion $\rm \frac{X_{Stage}}{X_{G}}$ with the feedback loop only (solid orange curve), with the feedback loop and sensor correction on (solid brown curve) and with feedback, sensor correction and feedforward on (solid black curve). The motion associated with sensor noise is not represented on this figure for clarity. The open loop (not represented) has a 30Hz unity gain frequency. The Sensor correction filter (dash cyan curve) is designed to provide extra isolation between 50mHz and 200mHz, whereas the feed forward filter (dash purple curve) provides isolation at 1Hz and above. At low frequencies where the loop gain is effectively infinite, the performance is limited by the low pass filter (dash green curve), and limited by the finite loop gain at higher frequencies. Typically, the low pass filter is tuned to provide as much isolation as possible in the control bandwidth at the cost of some gain peaking around the blend frequency (in this case $\sim$ 45mHz). Some sharp notches are also present in this filter to target known payload resonances. 

\section{Nominal Filters}

\begin{figure}[t]
\hspace*{-1.5cm}
\centering
\includegraphics[width=4in]{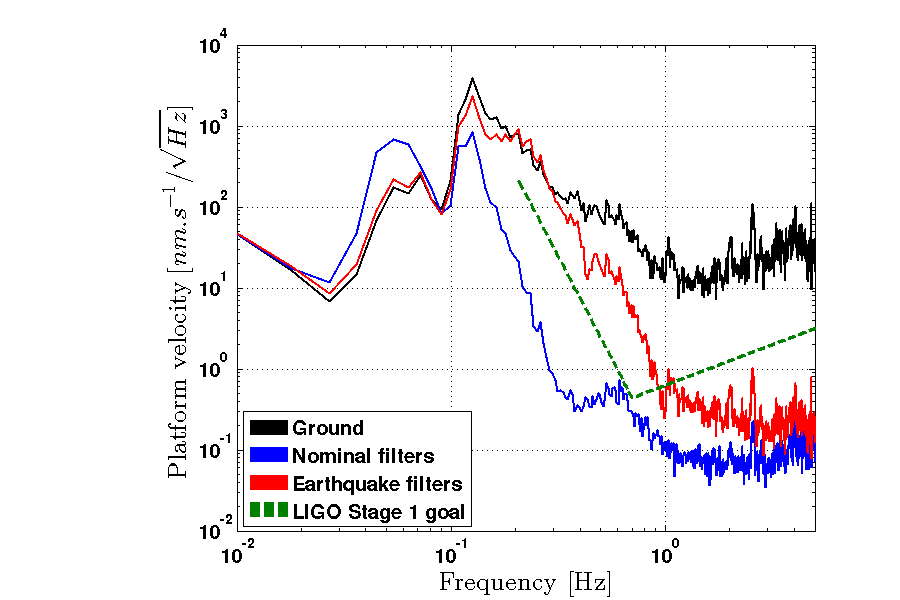}
\caption{Seismic isolation provided by BSC-ISI Stage 1 in the horizontal Y-direction at H1. The black curve represents a typical ground motion, and the blue curve the measured motion of the stage. The red curve is a simulation of the stage motion with the Earthquake filters in place. The dotted curve indicates the LIGO goal to obtain from 200mHz to higher frequencies for Stage 1.}
 \label{fig:ref}
\end{figure}

The nominal filters are designed to meet LIGO requirements and maximize the seismic isolation above 100mHz, at the expense of some gain peaking at lower frequencies. The blue curve in figure \ref{fig:ref} demonstrates the performance of this configuration at H1 during a typical ground motion period. The stage provides a factor 38 of isolation at 200mHz at the expense of a gain peaking of 4 at 54mHz. While this gain peaking is not a limitation during typical ground behavior, it becomes problematic during an earthquake.

Figure \ref{fig:bins} compares the performance of the stage during earthquakes. It shows a clear amplification of the ground motion by the stage over the bandwidth of interest (figure \ref{fig:bins}a versus figure \ref{fig:bins}b). With this configuration, the interferometer has a 50\% chance of losing lock for a stage velocity higher than 1000 nm/s.

\section{Earthquake Filters}

The goal of the earthquake control configuration is to find a good balance between limited gain peaking at low frequencies (because the nominal gain peaking is at the maximum earthquake power) and performance at higher frequencies. The gain peaking observed is directly dependent on the low-pass filter $\rm L_{disp}$ and the sensor correction filter $\rm H_{SC}$, as demonstrated in equation \ref{eq:gain}. The feedback and feedforward controllers stay untouched.

\begin{equation}
\begin{aligned}
\lim_{CP_{Stage}\to\infty} |X_{Stage}|^2={} & |(L_{disp}(H_{SC}-1)X_{G})|^{2} \\
      & + |H_{in}N_{in}|^{2} \\
      & + |L_{disp}N_{pos}|^{2}
\end{aligned}
\label{eq:gain}
\end{equation}

\begin{figure}[t]
\hspace*{-1cm}
\centering
\includegraphics[width=3.75in]{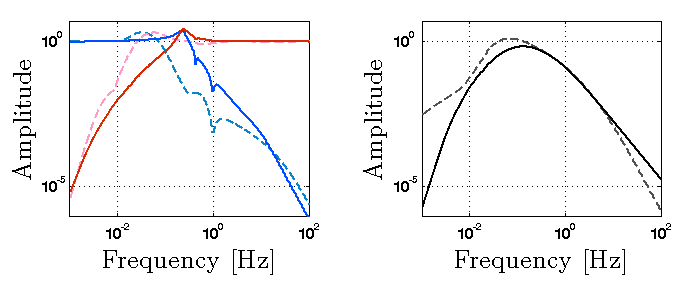}
\caption{Comparison of the filters used during O1 (dash lines) and the new designed filters for Earthquakes (solid lines). The left part of the figure show the complementary low-pass and high-pass filters. The right part shows the sensor correction filters.}
 \label{fig:new_filt}
\end{figure}

Figure \ref{fig:new_filt} shows a comparison between the newly designed low, high and sensor correction filters with the nominal filters. To move the gain peaking out of the earthquake band, the blend frequency between the low-pass and the high-pass filter has been increased from 45mHz to 250mHz, and the sensor correction has been slightly modified to be less aggressive. The overall gain peaking is reduced by a factor of 3.1 with this configuration. However, there is no longer any isolation at 200mHz, which greatly impacts the sensitivity of the detectors. The stage does not provide the required seismic isolation to observe gravitational waves when in operation, but will deliver enough isolation to keep the interferometer locked most of the time. It will not provide enough isolation to keep the detector locked if other disturbances, such as high-wind or distant storms, increase the ground motion around 200mHz.  Figure \ref{fig:ref} shows the simulated stage 1 motion with the new earthquake configuration (red curve).

\section{Improvement of LIGO robustness and duty cycle}

\begin{figure}[t]
\hspace*{-1.5cm}
\centering
\includegraphics[width=3.75in]{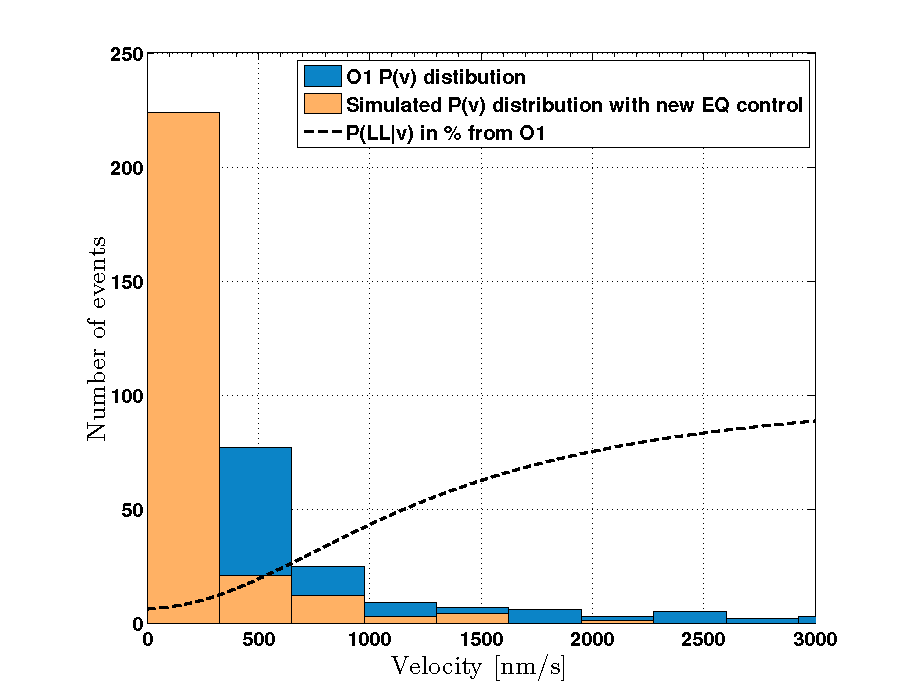}
\caption{New $\rm P(v_{new})$ distribution based on O1 data with $\rm P(v_{new})=\frac{P(v)}{1.5}$. As a reminder from figure \ref{fig:bins}, we plotted the probability of loosing lock as a function of stage velocity $\rm P(LL|v)$ in black.}
 \label{fig:pv}
\end{figure}

In this section we estimate the improvement in duty cycle due to reducing the sensitivity to earthquakes. Earthquake data collecting during O1 is used to simulate the effect of the new earthquake filters on Stage 1 velocity. We compare the averaged gain peaking induced by O1 earthquakes in the [30mHz-100mHz] band between the nominal and the earthquake filters. The velocity distribution will change from $\rm P(v)$ to $\rm P(v_{new})$, with $\rm P(v_{new})=\frac{P(v)}{1.5}$. Based on this new distribution (plotted in figure \ref{fig:pv}), we can re-calculate the probability of losing resonance $\rm P(LL)$ as written in equation \ref{eq:stats2}. We expect a 32.3\% reduction.

\begin{equation}
P(LL) = \sum_{\substack{velocity\\ bin}} P(LL|v)P(v_{new})
\label{eq:stats2}
\end{equation}

\begin{figure}[t]
\hspace*{-1.25cm}
\centering
\includegraphics[width=4in]{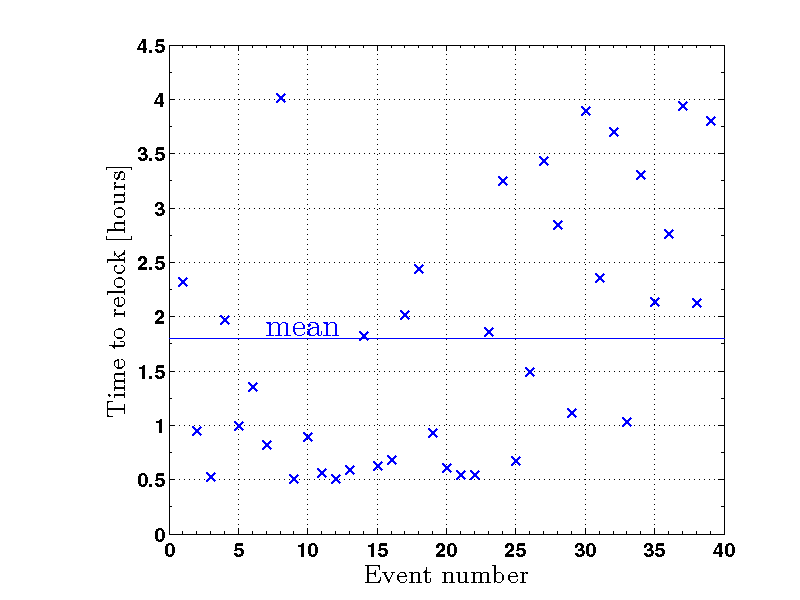}
\caption{Time spent returning to science mode after a problematic earthquake. If the time-span to re-acquire resonance was longer than 5 hours, it was assumed that the interferometer was put in stand-by and the event was not taken in account for the statistics. It takes 1.8 hours on average to return the instrument to its previous state.}
 \label{fig:timelock}
\end{figure}

Each time the interferometer goes out of operation there is an associated downtime. Due to the complexity of the instrument and the duration of an earthquake, it can take several hours to re-acquire resonance. During O1, it took an average of 1.8 hours to go back to science mode after an earthquake, as explained in figure \ref{fig:timelock}: H1 was down more than 111 hours because of earthquakes. If the number of times the interferometer goes out of resonance is reduced as expected, the downtime will be also be reduced from 111 to 75 hours.

\section{Implementation at the sites}

As the earthquake control configuration described in this paper doesn't meet LIGO observing requirements, it is important to switch to this configuration only when an problematic earthquake arrives at the site. An early alert system called Seismon has been developed and is currently installed at the observatories \cite{coughlin2017limiting}. Seismon predicts the arrival time of the primary, secondary and surface waves of an earthquake with a few seconds accuracy, and with a notification latency of less than 10 minutes. This prediction is using data from a world-wide network of seismic observatories that has been compiled by the United States
Geological Service (USGS) and can be accessed on the World-Wide Web \cite{usgs}. This gives enough time to switch the seismic configuration as needed. 

The second role of Seismon is to predict, based on a logistic regression algorithm explained in \cite{coughlin2017limiting}, the likelihood of the interferometer to go out of science mode. This gives us the information to assess if an upcoming event would be problematic. If there is a high likelihood of losing resonance, we would trigger a change in the control configuration using the LIGO automation system called Guardian.

Each interferometer is supervised by a state machine called Guardian \citep{rollins2016distributed}. It consists of state machine automation nodes capable of handling control changes automatically. It is composed of multiple nodes, organised in a hierarchical fashion for each system and subsystem. In the case of the BSC-ISI, multiple intermediate states are required to bring the platform from the initial state to a fully isolated chamber. More states can be added to enable the switch between low-pass, high-pass and sensor correction filters. This change of state will only require a few minutes, as it is possible to switch these filers in either direction without having to turn off the isolation loops. This is due to a filter switching system which is already part of the LIGO infrastructure \cite{blendswitch}.

\section{Conclusion}

In this paper, we have discussed the problem induced by earthquakes on the LIGO gravitational-wave detectors and presented a seismic control strategy to minimize their impact. This new design represents a 32.3\% improvement of the interferometers' robustness to earthquakes, which implies a direct increase in the number of detections for LIGO. We have described that the LIGO automation infrastructure is capable of switching to this new configuration based on early-warning predictions. Further effort will be spent on implementing this strategy for future observation runs.

\hspace{25cm}
\section{Acknowledgements}
The authors thank to the LIGO Scientific Collaboration for access to the data and gratefully acknowledge the support of the United States National Science Foundation (NSF) for the construction and operation of the LIGO Laboratory and Advanced LIGO as well as the Science and Technology Facilities Council (STFC) of the United Kingdom, and the Max-Planck-Society (MPS) for support of the construction of Advanced LIGO. Additional support for Advanced LIGO was provided by the Australian Research Council.
\bibliographystyle{unsrt}
\bibliography{reference_papers}{}

\begin{thebibliography}{10}

\bibitem{abbott2016gw150914}
BP~Abbott, R~Abbott, TD~Abbott, MR~Abernathy, F~Acernese, K~Ackley, C~Adams,
  T~Adams, P~Addesso, RX~Adhikari, et~al.
\newblock Gw150914: The advanced ligo detectors in the era of first
  discoveries.
\newblock {\em Physical review letters}, 116(13):131103, 2016.

\bibitem{aasi2015advanced}
J~Aasi, BP~Abbott, R~Abbott, T~Abbott, MR~Abernathy, K~Ackley, C~Adams,
  T~Adams, P~Addesso, RX~Adhikari, et~al.
\newblock Advanced ligo.
\newblock {\em Classical and Quantum Gravity}, 32(7):074001, 2015.

\bibitem{matichard2015seismic}
F~Matichard, B~Lantz, R~Mittleman, K~Mason, J~Kissel, B~Abbott, S~Biscans,
  J~McIver, R~Abbott, S~Abbott, et~al.
\newblock Seismic isolation of advanced ligo: Review of strategy,
  instrumentation and performance.
\newblock {\em Classical and Quantum Gravity}, 32(18):185003, 2015.

\bibitem{abbott2002seismic}
R~Abbott, R~Adhikari, G~Allen, S~Cowley, E~Daw, D~DeBra, J~Giaime, G~Hammond,
  M~Hammond, C~Hardham, et~al.
\newblock Seismic isolation for advanced ligo.
\newblock {\em Classical and Quantum Gravity}, 19(7):1591, 2002.

\bibitem{abbott2004seismic}
R~Abbott, R~Adhikari, G~Allen, D~Baglino, C~Campbell, D~Coyne, E~Daw, D~DeBra,
  J~Faludi, P~Fritschel, et~al.
\newblock Seismic isolation enhancements for initial and advanced ligo.
\newblock {\em Classical and Quantum Gravity}, 21(5):S915, 2004.

\bibitem{senscor}
B~Lantz.
\newblock {\em Description of the Sensor Correction FIR and IIR Filter
  Components}, 2012.
\newblock \url{https://dcc.ligo.org/LIGO-T1200285/public}.

\bibitem{coughlin2017limiting}
M~Coughlin, P~Earle, J~Harms, S~Biscans, C~Buchanan, E~Coughlin, F~Donovan,
  J~Fee, H~Gabbard, M~Guy, et~al.
\newblock Limiting the effects of earthquakes on gravitational-wave
  interferometers.
\newblock {\em Classical and Quantum Gravity}, 34(4):044004, 2017.

\bibitem{usgs}
{\em Data from the IRIS network of the USGS can be found at}.
\newblock \url{https://earthquake.usgs.gov/earthquakes/}.

\bibitem{rollins2016distributed}
JG~Rollins.
\newblock Distributed state machine supervision for long-baseline
  gravitational-wave detectors.
\newblock {\em arXiv preprint arXiv:1604.01456}, 2016.

\bibitem{blendswitch}
R~Kurdyumov, C~Kucharczyk, and B~Lantz.
\newblock {\em Blend Switching User Guide}, 2012.
\newblock \url{https://dcc.ligo.org/LIGO-T1200126/public}.

\end{thebibliography}

\end{document}